

\magnification=\magstep1
\baselineskip 17pt plus 2pt

\font\headl=cmfib8 scaled \magstep3


\newcount\refnumber
\refnumber=0
\immediate
\openout 1 = ref
\def\ref#1{\global\advance\refnumber by 1
            \immediate\write 1{\string\item{\number\refnumber )}\string\lz#1}}



\newcount\equnumbera
\equnumbera=0
\def\eqna{\global\advance\equnumbera by 1
         \eqno(\the\equnumbera)}


\def\rama{\vadjust{\vbox to 0pt
         {\vss
          \hbox to \hsize{\hskip\hsize
                          \quad
                          $\Leftarrow$\hss}
          \vskip3.5pt}}}




\def\xvi{x_i}
\def\pvi{p_i}
\def\argxp{\xvi ,\pvi }
\def\resqq{ \chi _{qq} (\omega ) }
\def\resppqq{ \chi _{qq}^{\prime \prime } (\omega ) }

\def\fqpt{f(q,P,t)}
\def\varpib{\varpi_b}
\def\varpia{\varpi_a}
\def\etab{\eta_b}
\def\om{\omega}
\def\absc{\mid C_b \mid }

\def\kwigner{k^{\rm W} (\beta ;Q,P)}
\def\kglob{k_{\rm glob}^{\rm W} (Q,P)}

\def\ing{A}
\def\errpar{-\invsigmaqq \ing^2 -\invsigmapp}
\def\eqsigmaqq{{\textstyle \sum }_{qq} }
\def\eqsigmapp{{\textstyle \sum }_{pp} }
\def\eqsigmaqp{{\textstyle \sum }_{qp} }
\def\eqsigmaqqa{{\textstyle \sum }^{a}_{qq} }
\def\eqsigmappa{{\textstyle \sum }^{a}_{pp} }
\def\eqsigmamn{{\textstyle \sum }_{\mu \nu}}
\def\invsigmaqq{{\textstyle \sum }^{b}_{qq} }
\def\invsigmapp{{\textstyle \sum }^{b}_{pp} }

\def\prefib{\zeta_b}
\def\prefia{\zeta_a}
\def\toth{\hat {\cal H}}
\def\zdampglob{{\cal Z}_{\rm glob}}
\def\zdamp{{\cal Z}}
\def\zdampa{{\cal Z}_a}
\def\zdampinv{{\cal Z}_b}
\def\zdampinvabs{\vert \zdampinv \vert}
\def\hbomt{\left({\hbar\varpib \over 2T}\right)}

\def\qucof{{\rm f}_Q}

\def\ratekram{$$ R_K ={\varpi_a \over 2\pi}
     \left(\sqrt{1+ \etab^2} - \etab  \right) e^{-\beta B}
    \eqna $$}

\def\htotsb{$$ \toth(\argxp,Q,P) = \hat H_{\rm B}(\argxp) +\hat H_{\rm
    SB} (\argxp,Q,P) + \hat H_{\rm S}(Q,P)       \eqna $$}
\def\defrate{$$ R \equiv {j_b\over N_a} =
    {\int_{-\infty}^\infty \,{\rm d}P \;\;{P\over M} \;
    k_{\rm glob}^{\rm W} (\beta ;Q\sim Q_b=0,P) \over
    \int_{Q_a-\Delta}^{Q_a+\Delta}\,{\rm d}Q\,\int_{-\infty} ^\infty \,{\rm d}P
 \;\;
    k_{\rm glob}^{\rm W} (\beta ;Q,P)}
    \eqna $$}
\def\traneq{$${\partial  \over \partial t} \fqpt = \Biggl[
    -{\partial\over\partial q}{P\over M} + {\partial\over\partial P}Cq
    +{\partial\over\partial P}{P\over M}\gamma +
     D_{qp}{\partial^2\over\partial q\partial P} + D_{pp}
    {\partial^2\over\partial P\partial P} \Biggr] \fqpt \eqna $$}
\def\eqaverqq{$$ \eqsigmaqq =\int \! {{\rm d}\omega \over 2\pi }\,
  \coth \left( {\omega \over 2T}\right) \, \resppqq \eqna $$}
\def\distmin{$$ k_{\rm osc}^{\rm W} (q,P)=
     \prefia {e^{\beta B}}
    \exp\Bigl(-{P^2 \over 2 \eqsigmappa } -{(q-q_a)^2 \over 2
     \eqsigmaqqa}\Bigr)
    \eqna $$}
\def\transfomega{$$\varpia =\sqrt{{C_a\over M_a}} \quad
    \Longrightarrow \quad i\varpib =\sqrt{{-\vert C_b \vert \over M_b}}
    \eqna $$}
\def\stationingolddec{$$ k_{\rm glob}^{\rm W} (Q\sim Q_b,P) \sim
       k_{\rm I}^{\rm W}(q,P) =\prefib \exp\Bigl(
     -{P^2 \over 2 \invsigmapp} -{q^2 \over 2 \invsigmaqq}\Bigr)
    \int_{-\infty}^{P-\ing q}\;{{\rm d}\,u\over \sqrt{2\pi \sigma}}
    \;\exp\Bigl[-{u^2\over 2\sigma}\Bigr]    \eqna $$}
\def\resrate{$$ R ={\zdampinvabs \over \zdamp _a}
    {\varpib\over 2\pi} \left(\sqrt{1+ \etab^2} - \etab\right)e^{-\beta B}
    \equiv \qucof R_K   \eqna $$}
\def\partfunceig{$$ \zdampa= {\cal N}
    \left(\prod_{n=-\infty}^{+\infty} {\lambda_n^a \over M_a}\right)^{-1/2}
    \eqna$$}

\def\quantcorr{$$ \qucof = \prod_{n=1}^\infty
    {\nu_n^2 + \nu_n \Gamma_a(\nu_n) + \varpia^2
    \over
    \nu_n^2 + \nu_n \Gamma_b(\nu_n) - \varpib^2}
    \eqna $$}

\def\quavirial{$$ \langle C_bq^2/2\rangle_b = \langle P^2/2M_b\rangle_b =
    {1\over 2}T^*(i\varpib) = {1\over 2}{\hbar \varpib \over 2}
    \cot\hbomt \eqna $$}

\def\lzhofrep{H. Hofmann, to be published in Phys. Rep.}

\def\lzhoso{H. Hofmann and R. Sollacher, Ann. Phys. 184 (1988) 62}

\def\lzhosaoc{H. Hofmann, R. Samhammer and G. Ockenfu\ss , Nucl. Phys. A496
     (1989) 269}

\def\lzhofkitse{H. Hofmann, D. Kiderlen and I. Tsekhmistrenko, Z.Phys.A
    341 (1992) 181}

\def\lzcalleg{A.O. Caldeira and A.J. Leggett, Ann. Phys. 149 (1983) 374}

\def\lzkram{ H. A. Kramers, Physica 7 (1940) 284}

\def\lzingold{G.-L. Ingold, Thesis, Univ. Stuttgart 1988}
\def\lzankgraing{J. Ankerhold, H. Grabert and G.-L. Ingold, to be published}

\def\lzhofing{H. Hofmann and G.-L. Ingold, Phys. Lett. 264B (1991) 253}

\def\lzgraschring{H. Grabert, P. Schramm and G.-L. Ingold, Phys. Rep.
     168 (1988) 115}
\def\lzgraolwei{H. Grabert, P. Olschowski and U. Weiss, Phys. Rev.B 36
     (1987) 1931}
\def\lzgraweitalk{H. Grabert, U. Weiss and P. Talkner, Z.Physik B 55
     (1984) 87}
\def\lzankergra{J. Ankerhold and H. Grabert, Physica A 188 (1992) 568}

\def\lzrisehangweis{P.S. Riseborough, P. H\"anggi and U. Weiss,
    Phys. Rev. A31 (1985) 471}
\def\lzeckernet{U. Eckern et.al., J. Stat. Phys. 59 (1990) 885}
\def\lzhantalbor{P. H\"anggi, P. Talkner and M. Borkovec, Rev. Mod. Phys. 62
    (1990) 251}

\def\lznixscheuvaut{J.R. Nix, A.J. Sierk, H. Hofmann, F. Scheuter,
     D. Vautherin, \hfill \break Nucl. Phys. A424 (1984) 239}

\def\lzfroebtill{P. Fr\"obrich and G.-R. Tillack, Nucl. Phys. A540
     (1992) 353}

\topskip=2truecm
\centerline{{\headl Fission decay rates}}
\medskip
\centerline{{\headl determined from a quantal transport equation}}
\bigskip
\bigskip
\bigskip
\centerline{by}
\bigskip
\centerline{\bf Helmut Hofmann}
\centerline{Physik-Department, TU M\"unchen}
\centerline{James Franckstra\ss e, D-85747 Garching}
\bigskip
\centerline{\bf Gert-Ludwig Ingold}
\centerline{Fachbereich Physik, Universit\"at-GH Essen}
\centerline{D-45117 Essen 1}
\bigskip
\centerline{and}
\bigskip
\centerline{\bf Markus H. Thoma}
\centerline{Institut f\"ur Theoretische Physik, Universit\"at Giessen}
\centerline{D-35392 Giessen}
\vskip 1.5cm
\parindent=0pt
The decay of a metastable system is described by extending
Kramers' method to the quantal regime. For temperatures above twice the
crossover value we recover the result known from applying Euclidean
path integrals to solvable models. Our derivation is not restricted to
a linearly coupled heat bath of oscillators, and thus applicable to
nuclear systems.

\vskip 1cm
\vfill
\eject
\topskip=10pt

\parindent=20pt
\ref{kram}
\edef\nkram{\the\refnumber}
In his famous paper Kramers [\nkram ] derived a formula which took
into account the implications of dynamics on the decay rate of a
metastable system. Applied to the situation portrayed in Fig.1 it can
be written as: 
\ratekram
\edef\nratekram{\the\equnumbera}
Here, $\beta = (kT)^{-1}$, $B$ is the barrier height and $\varpia $
the frequency of the  motion around the potential minimum (fulfilling
the relation $M_a\varpia ^2 = C_a=\partial^2V(Q)/\partial Q^2 \vert _a$
with $M_a$ being the inertia).  The $\etab  = \gamma_b /
(2\sqrt{M_b\absc})$ measures the effective damping rate, calculated at
the barrier from the friction coefficient $\gamma_b$, the inertia $M_b$
and the stiffness coefficient $C_b=\partial^2V(Q)/\partial Q^2 \vert
_b$.

Expression (\nratekram) is valid for damping rates $\etab$
not smaller than about $0. 2$. Furthermore, one has to assume
the barrier big enough to keep the decay rate $R$ sufficiently
small. Under such conditions the process can be viewed as
quasi-stationary with the outward flux $j_b$ at $Q_b$ being
constant for macroscopically large times. The decay rate $R$ is then
simply given by the ratio of the current divided by the number of
particles $N_a$ caught in the well near $Q_a$, $R =j_b/N_a$.

\ref{calleg}
\edef\ncalleg{\the\refnumber}
\ref{hantalbor}
\edef\nhantalbor{\the\refnumber}
\ref{graolwei}
\edef\ngraolwei{\the\refnumber}
\ref{ingold}
\edef\ningold{\the\refnumber}
\ref{ankgraing}
\edef\nankgraing{\the\refnumber}
\ref{eckernet}
\edef\neckernet{\the\refnumber}
It has always been a challenge to extend Kramers' result to the quantal
regime. However, it was only after one had learned to apply the
instanton trick to this problem that a decent solution could be found.
Among the vast literature on this subject of "dissipative tunneling" we
would only like to refer to the classic paper [\ncalleg], to
[\nhantalbor] as a general recent review and to [\ngraolwei] as a reference
to formal details. Common to all these approaches is the application of
the technique of path integrals for imaginary time propagation. This is
feasible if for the basic Hamiltonian one assumes a form like:
\htotsb
\edef\nhtotsb{\the\equnumbera}
in which two restrictions are imposed. First, the coupling  is assumed
to be of the form $H_{\rm SB}=QF$ with $F=\sum_i c_i\xvi$.
Secondly, the heat bath must be represented by a set of oscillators. It
is needless to stress that some measures have to be taken in order to
render the final equations of motion irreversible.

A first major step to unify these results with the usual concepts of
transport theory has been undertaken in [\ningold, \nankgraing]. Still using
 path
integrals, the author has been able to obtain from real time
propagation a quantal version of Kramers' stationary solution for the inverted
 oscillator,
which he could then use to rederive the form of the known quantal correction
factor. This derivation is valid above a critical temperature $T_c=2 T_0$ with
$T_0$ being the so called crossover value [\nhantalbor, \ngraolwei]. We
will come back to the physical significance of this restriction.

A full-fledged transport theory has been formulated in [\neckernet] within
a quasi-classical approach based on a quantal Langevin equation.
Unfortunately, the authors have not been able to establish a direct
connection to the results mentioned above. This will be the main goal
of the discussion to come. We will proceed by exhibiting first the
formal details, and then discuss later both the physical conditions
of the derivation as well as consequences for possible applications.

Let us begin by formulating the decay rate in terms of Wigner
functions $\kwigner$. Borrowing from the ideas of Kramers we may write:
\defrate
\edef\ndefrate{\the\equnumbera}
Here $\kglob$ is meant to represent the full global solution to our problem.
Under the conditions mentioned already above, and for not too small
temperatures (see below), this global function can be approximated by
local ones valid at the barrier or at the potential minimum,
respectively. They are obtained by appropriate linearizations.
Trivially an overall normalization factor drops out of eq.(\ndefrate)
and it is almost obvious that the local solutions should not be
normalized in themselves. In this context it is useful, for a moment,
to think of a stationary situation. Then $\kglob$ would simply be the
Wigner transform of $\hat k_{\rm glob}=\exp(-\beta
\toth) $. As a consequence of this transformation one gets
a prefactor (to the exponential factor depending on the coordinates and
momenta),  which in the end may be found from the partition function
$\zdampglob$. The linearization procedure we spoke of will in such a
case have to be performed on $\toth $. In this way it will lead to both
a local partition function as well as to a local prefactor $\zeta$.
Both will depend on the local frequency rendering obvious that the
$\zeta$ will have to be different at the well and at the barrier. Our
situation differs from the one just described only in the sense that we
have to deal with a dynamical situation, albeit in quasi stationary
approximation, whose solutions must be found from transport equations.

\ref{hosaoc}
\edef\nhosaoc{\the\refnumber}

\ref{hoso}
\edef\nhoso{\the\refnumber}
\ref{hofrep}
\edef\nhofrep{\the\refnumber}

In [\nhosaoc] a Hamiltonian of type (\nhtotsb) has been used to derive
a transport equation for local collective motion.  This Hamiltonian can
be obtained self-consistently by applying an appropriate linearization
procedure with respect to the macroscopic degrees of freedom (see
[\nhoso]).  No restrictions need to be imposed on the "bath"
Hamiltonian $\hat H_{\rm B}$ and the $F$ appearing as a factor in $\hat
H_{\rm SB}$ (see text below eq.(\nhtotsb)). As we shall see, for the
present purpose it suffices to expand around values $Q_0$ which
correspond to extremal points of the effective potential (for the
general case see eg.  [\nhosaoc] or [\nhofrep]). Then the result can be
written as: 
\traneq
\edef\ntraneq{\the\equnumbera}
with $q=Q-Q_0$. (With respect to $P$ we assume from the start that our
Hamiltonian is at most quadratic).
Besides the transport coefficients for average (local)  motion, there
appear the two diffusion coefficients $D_{pp} = (\gamma / M)\eqsigmapp$
and $D_{qp} = C \eqsigmaqq - (1/ M)\eqsigmapp $, with
the equilibrium fluctuation $\eqsigmaqq$
given by the fluctuation dissipation theorem (FDT):
\eqaverqq
\edef\neqaverqq{\the\equnumbera}
Similar expressions hold for $\eqsigmapp$ and $\eqsigmaqp$ if for each
appearance of the kinetic momentum $P$ a factor $M\om $ is multiplied to
the integrand. As a consequence, the $\eqsigmaqp$ vanishes because of a
simple symmetry, which is the reason for $D_{qq}=0$.

\ref{graweitalk}
\edef\ngraweitalk{\the\refnumber}
\ref{risehangweis}
\edef\nrisehangweis{\the\refnumber}

The $\resppqq$ is the dissipative part of the susceptibility $\resqq$
which measures the response of $\langle q \rangle_\om$ to an external
 "field" $q_{\rm ext}(\om)$ in linear order. These functions
contain information about all possible modes of our system, as given by
the Hamiltonian (\nhtotsb). Clearly, we are interested only in the
collective ones, namely those whose average behaviour we have
parametrized in terms of the transport coefficients $M, \gamma$ and
$C$, defined for local motion. To render the dynamics of the
fluctuations consistent, we have to
calculate the diffusion coefficients from the corresponding
approximations. That means to replace in (\neqaverqq) the locally valid
$\resppqq$ by the dissipative part of the oscillator response function being
 defined
by: $ \chi_{\rm osc}^{-1}(\om ) q(\om ) \equiv ( -\om ^2 M -i\om \gamma
+ C) q(\om )=-q_{\rm ext}(\om )$.  We are aware that this restriction
will finally cause the expected problem of a diverging
$\eqsigmapp$. To regularize the integral appearing there one has to
introduce one other constant.  For example, this could be done by way
of a cut off [\nhosaoc] or through a Drude regularization
(cf.[\ngraweitalk] and [\nrisehangweis]). For our present purpose this
problem is not really relevant.

We are ready now to evaluate the decay rate. Let us look first at the
denominator of (\ndefrate). The linearization we spoke of before
 means to put $ k_{\rm glob}^{\rm W} (Q\sim Q_a,P) \sim k_{\rm osc}^{\rm W}
(q,P)$ and to use the following stationary solution of (\ntraneq):
\distmin
\edef\ndistmin{\the\equnumbera}
with the prefactor given as $2\pi \zeta_a = \zdampa / \sqrt{\eqsigmaqqa
\eqsigmappa }$, which normalizes (6) to $Z_a$ as discussed above.
(Here and in the following the $\zdamp $'s always refer
to the effective macroscopic part of the partition function). In order
to calculate the number of particles $N_a$ we may replace $\Delta$ by
$\infty$, provided we have $\vert Q_a\vert \gg \sqrt{\eqsigmaqqa}$.
\ref{hofing}
\edef\nhofing{\the\refnumber}
Next we turn to the barrier region. Here the stiffness is negative and
the basic expressions for the diffusion coefficients become somewhat
delicate, as the "equilibrium fluctuations" loose their immediate
physical meaning. Nevertheless, all formulas can be continued
analytically by replacing
\transfomega
\edef\ntransfomega{\the\equnumbera}
everywhere. In this way the $\eqsigmaqq$ becomes negative, as already
known from Kramers' case of high temperatures.
For the prefactor at $Q_b$ this implies to write
$2\pi \zeta_b= \zdampinv / \sqrt{\invsigmaqq \invsigmapp }
=\zdampinvabs / \sqrt{-\invsigmaqq\invsigmapp}$.

As for the functional dependence on $Q$ and $P$, a form like the one obtained
from (\ndistmin) still solves our equation of motion (\ntraneq), now
written for the local motion around $Q_b=0$. However, this would not be
a decent approximation to $\kglob$. Indeed, since the barrier is
unbound to the right we expect a solution with a finite current. Such a
solution has been displayed in [\nhofing] and applied to
computations of interest in nuclear fission, but not to the decay rate.
Its form had been found before in [\ningold] within a path integral
description. In our present notation it reads 
\stationingolddec
\edef\nstationingolddec{\the\equnumbera}
with $\sigma = \errpar $ and
$M_b\varpib (\sqrt{1+ \etab^2} - \etab )  A = \invsigmapp / (-\invsigmaqq) $

The final expression for the decay rate is easily obtained. After
putting into (\ndefrate) both (\ndistmin) and (\nstationingolddec) the
integrals can be calculated by elementary methods to give:
\resrate
\edef\nresrate{\the\equnumbera}
As observed already in [\ningold], the quantum correction factor
$\qucof$ can be expressed by the absolute value of the ratio of the local
partition functions at $a$ and $b$. In the remaining part of our letter
we will discuss the physical relevance of our result together with a
critique of the present derivation.

\ref{froebtill}
\edef\nfroebtill{\the\refnumber}
1) {\it Comparison with previous derivations}.
The partition functions appearing in (\nresrate) are readily calculated
for the model of the bilinear coupling to a set of oscillators, for
example with path integrals (cf. e.g.[\ngraolwei] and our discussion below
eq.(\nhtotsb)).  Again take first the case of the potential minimum
where one gets:
\partfunceig
\edef\npartfunceig{\the\equnumbera}
with the eigenvalues
$\lambda_n^a = M_a\nu_n^2 + \vert\nu_n\vert \gamma_a(\vert \nu_n\vert )
+ M_a\varpia^2 $. Here, $\nu_n = 2n \pi / \hbar \beta$
and ${\cal N}$ is the usual reference factor being independent of
$\varpia, M_a$ and $\gamma_a$. (We follow the notation of [\ngraolwei]
allowing frequency dependent friction coefficients $\gamma(\om)$
for an eventual Drude regularization).  The $\zdampinv$ is obtained from the
transformation (\ntransfomega).  We observe that the mode $n=0$ has a
negative eigenvalue, which for this model proves the conjecture about
$\zdampinv$ stated above, namely that $\zdampinv$
becomes purely imaginary. For temperatures above $T_0$, with the
corresponding $\beta_0$ being a solution of
$\lambda_1^b=\lambda_{-1}^b=0$, all other eigenvalues are positive.
For the quantum correction factor one thus obtains:
\quantcorr
\edef\nquantcorr{\the\equnumbera}
with $\Gamma = \gamma / M$.
This expression has been derived before
within the formulation of "dissipative tunneling", but for the case
$M_a=M_b$ and $\gamma _a=\gamma _b$
\footnote{*}{The case of variable inertia and friction has been treated
in [\nfroebtill] within a phenomenological ansatz; see our discussion to
come in 4).}.  We present it in Fig.2 for a whole range of effective
damping factors $\etab$. At the crossover temperature $T_0$ this
formula diverges, but it can be regularized after considering
deviations of third order from the saddle points [\ngraolwei].  As we
shall see now our derivation is restricted to temperatures of
approximately twice that value, $T >2 T_0$.

\ref{ankergra}
\edef\nankergra{\the\refnumber}

2) {\it The limits of our approach}.
In our discussion above we have been very sloppy about any convergence
problem appearing in the integrals. Looking back at formula (\ndefrate)
we realize that problems may come from the numerator of this
expression. We certainly need a positive $\invsigmapp $. However, the
more crucial quantity turns out to be the $\invsigmaqq$. In
Kramers'case it is known to be negative such that the Gaussian factor
in (\nstationingolddec) increases with increasing $q$.
To study these expressions in general and for sizable friction is
somewhat elaborate, at least for the regime where quantal effects are
important.  For a first orientation, let us therefore look at weakly
damped motion, to postpone a discussion of the general case to future
publications (see e.g.[\nhofrep]).  Evaluating the $\eqsigmamn$ from
(\neqaverqq) to zero order in $\gamma$ one gets: 
\quavirial
\edef\nquavirial{\the\equnumbera}
which is nothing else but the quantal virial theorem continued
analytically to the imaginary frequency of the barrier, making apparent
the fact that the "equilibrium fluctuation" in $q$ is negative at the
barrier. (To get Kramers' case of high temperature we just have to
replace $T^*$  by $T$).  For a situation like the one behind
(\nquavirial) the condition we have to pose clearly is
$T>T_c=2T_0\equiv \hbar \varpib / \pi $. It guarantees that
$\eqsigmapp$ is positive and $\eqsigmaqq$ negative. Notice, that for
this situation it is easy to observe that also the $\sigma$ of
eq.(\nstationingolddec) is positive. At $T=T_c=2T_0$
the Gaussian factor appearing in (\nstationingolddec)  becomes
indifferent as function of $q$ and regularization procedures would have
to be employed [\nankergra ,\nankgraing].

\ref{graschring}
\edef\ngraschring{\the\refnumber}

\ref{nixscheuvaut}
\edef\nnixscheuvaut{\the\refnumber}

3) {\it On the quantum nature of this approach}.
Our result clearly demonstrates that the transport equation
(\ntraneq) contains quantum effects. It is clear that the latter are
there on a semiclassical level only, as global motion is treated in a
locally harmonic fashion. Nevertheless it is interesting to see the
effects contained in a differential transport equation.

Without any doubt this equation causes problems for the basic
uncertainty relations {\it at short times}. This is no surprise as we
have left out memory effects as well as the implications of initial
correlations. That both these effects must be there can be studied in the
oscillator model again. One may look, for instance, at the results of
[\ngraschring] for the time evolution of the second moments, say for
the case of a positive stiffness. Compared with the evolution obtained
from (\ntraneq), one sees that both agree for large times. This is to
be expected, of course, from the very construction of the transport
equation [\nhosaoc]:  The diffusion coefficients are chosen such as to
guarantee the correct equilibrium, as defined by the FDT.

The lesson we learn from our present results is that these features
prevail also for the case of the inverted oscillator. Again, this does
not come unexpected. One knows both from [\nnixscheuvaut] and
[\ningold] that for the inverted oscillator time dependent solutions
"relax" to stationary ones.  Apparently, in the regime of temperatures
we look upon here, it is the long time behaviour of the solution which
matters---and this behaviour contains the adequate quantum feature.
Further details will be discussed elsewhere, see e.g.[\nhofrep].

4) {\it Advantages of our approach}.
Although our method does not apply at very low temperatures, we
hope to contribute to the understanding of quantum effects
in dissipative dynamics. We see at least the following three
favorable circumstances.

\parindent=0pt
i) It should be of some theoretical interest to see effects of
"dissipative tunneling" come out of a transport equation proper.

ii) Our derivation is {\it not} based on the assumption of a {\it
linear coupling} to a {\it heat bath} which does {\it not change with
$Q$}. It is commonly understood that such a model would be quite
unrealistic for nuclear fission, a case discussed already by Kramers.
In a Hamiltonian like (\nhtotsb) the $\argxp $ represent the dynamics
of the nucleons.  For a first approximation there is no way around a
mean field approach, which poses important constraints both on the
Hamiltonian itself as well as on the treatment of the coupling (see
[\nhoso] and further literature cited there). First of all, $\hat
H_{\rm SB}$ {\it must} have a term involving the collective momentum
$P$. Secondly, the dependence of $\hat H_{\rm B}$ and $\hat H_{\rm SB}$
on the $\xvi$ cannot just be of first and second order. This feature
becomes already evident by looking at the simple but well examined
case of multipole vibrations around a potential minimum. It is only for
dipole excitations that the $\hat H_{\rm SB}$ can be taken linear in
the $\xvi$.  Thirdly, for large scale motion both $\hat H_{\rm B}$ as
well as $\hat H_{\rm SB}$ must vary along the collective path.  Within
the locally harmonic approach one is able to incorporate these effects
both for the construction of the Hamiltonian $\toth(\argxp,Q,P)$ itself
[\nhoso] as well as for the derivation of the equations of motion
[\nhosaoc].

We have mentioned already ref.[\nfroebtill]. There the effects of
variable inertia and friction on the decay rate have been taken into
account by the following procedure:  It was assumed that a Hamiltonian
like (\nhtotsb) exists for {\it global} motion, with a factorized form
for the $\hat H_{\rm SB}$; the latter was supposed just to depend on
the coordinates, thus neglecting self-consistency; some
phenomenological ansatz was made both for the form factor of $\hat
H_{\rm SB}$ as well as for the (unperturbed) inertia in $\hat H_{\rm
S}$,  neglecting the renormalization of the conservative forces through
the coupling; the collective variables were linearly coupled to a heat
bath of oscillators with fixed frequencies.

\ref{hofkitse}
\edef\nhofkitse{\the\refnumber}
iii) Our approach may allow for interesting practical applications,
even beyond the mere calculation of decay rates.  For instance, one is
not bound to look at cases where the temperature stays constant.
Moreover, as we look at real time propagation our transport equation
can easily be used for Monte Carlo simulations.  Notice, that our
locally harmonic approximation concurs ideally with the small time
steps employed in these calculations. Such simulations have proven
successfully in many fields of physics. Let us only allude to the
following examples where instabilities play a dominant role: once more
nuclear fission, chemical and surface reactions, nucleation and
spinodal decompositions.  The lesson we learn from the present study is
how simple it can be to incorporate quantum effects: Suppose one wants
to treat time evolution of the average values on the basis of classical
physics. The use of proper diffusion coefficients restores quantum
features on the level of the second moments, a procedure which is quite
easy to perform within the locally harmonic approximation. In the case
discussed here average dynamics is represented by our collective
variable $Q(t)$, but an extension to time evolution of the general mean
field is straightforward. For the latter the appropriate classical
means is the Landau-Vlasov equation. A linearization allows to deduce
response functions which in turn determine quantum statistical
fluctuations and hence generalized diffusion coefficients
(cf.[\nhofkitse]).  
\bigskip\bigskip

{\it Acknowledgments:} We would like to thank H. Grabert, P. Fr\"obrich
and R. Sollacher for fruitful discussions and valuable suggestions.
\bigskip\bigskip
\vfill
\eject
\leftline{{\it FIGURE CAPTIONS}}
\bigskip
Fig.1: A schematic plot of the effective potential.
\bigskip
Fig.2: The quantum correction factor $\qucof$ for the case of both
constant inertia $M_a=M_b$ and friction $\gamma_a=\gamma_b$, and  the
following values of the effective damping rate $\etab = 10, 5, 2, 1,
0.5$, from left to right, respectively. The arrows show
the corresponding crossover temperatures $T_0$.
\goodbreak
\bigskip
\leftline{{\it REFERENCES}}
\medskip
\parindent=20pt

\item{1)}\lzkram
\item{2)}\lzcalleg
\item{3)}\lzhantalbor
\item{4)}\lzgraolwei
\item{5)}\lzingold
\item{6)}\lzankgraing
\item{7)}\lzeckernet
\item{8)}\lzhosaoc
\item{9)}\lzhoso
\item{10)}\lzhofrep
\item{11)}\lzgraweitalk
\item{12)}\lzrisehangweis
\item{13)}\lzhofing
\item{14)}\lzfroebtill
\item{15)}\lzankergra
\item{16)}\lzgraschring
\item{17)}\lznixscheuvaut
\item{18)}\lzhofkitse

\vfill

\end